\documentclass[preprint,authoryear,12pt]{elsarticle}

\usepackage{epsfig}

\usepackage{amssymb}
\usepackage{subfig}             
\usepackage{amsmath}

\usepackage[ps2pdf,%
a4paper=true,%
breaklinks=true,%
colorlinks=true,%
pdfauthor={First Author et al.},%
pdftitle={Template for manuscripts in Advances in Space Research}%
]{hyperref}

\journal{Advances in Space Research}

\begin{document}

\begin{frontmatter}



\title{Object Image Linking of Earth Orbiting Objects in the Presence
of Cosmics}


\author{Carolin Fr{\"u}h\corref{cor}\fnref{footnote2}}
\author{Thomas Schildknecht\fnref{footnote2}}
\address{Astronomical Institute, University of Bern, Sidlerstrasse 5, 3012
  Bern, Switzerland}
\cortext[cor]{Corresponding author}
\fntext[footnote2]{Astronomical Institute, University of Bern, Sidlerstrasse 5, 3012
  Bern, Switzerland, phone: +41 76 4906 535, fax: +41 31 631 3869}
\ead{frueh@aiub.unibe.ch}


\ead{thomas.schildknecht@aiub.unibe.ch}

\author{This is a preprint, final version of this article is available at Advances in Space Research: http://dx.doi.org/10.1016/j.asr.2011.10.021}

\begin{abstract}
   In survey series of unknown Earth orbiting objects, no a
     priori orbital elements are available. In surveys of wide field
     telescopes possibly many non-resolved object images are present on the single frames
     of the series.
   Reliable methods have to be found to associate the object images stemming
     from the same object with each other, so-called linking.  The presence of cosmic ray events, so-called Cosmics,
   complicates reliable linking of non-resolved images. The tracklets of
   object images allow to extract exact positions for a first orbit
   determination.\\
\\
   A two step method is used and tested on observation frames of space debris
     surveys of the ESA Space Debris Telescope, located on Tenerife, Spain: In a first step a cosmic filter is applied in
     the single observation frames. Four different filter approaches are
     compared and tested in performance. In a second step, the detected object
   images are linked on observation series based on the assumption of a linear
   accelerated movement of the objects over the frame during the series, which
 is updated with every object image, that could be successfully linked.\\
\\
As results it has to be stated. Firstly, the automatic discrimination of cosmics and non-resolved object images on
     single frames remains difficult, especially for traces with a size of only a few
     pixel. Secondly, on observation series, the object image linking of GEO, GTO and
     MEO object images is reliable for a moderate amount of object images or
     cosmics on the frames if at least three object images are available per
     object and for four object images per object even in the presence of several hundred cosmics or
   images on the frames with a simple approach of an updated linear
   accelerated movement even over long observation series.
\end{abstract}


\end{frontmatter}

\parindent=0.5 cm

\section{Introduction}

The Astronomical Institute of the University of Bern (AIUB) performs
surveys for unknown space debris object in geostationary (GEO),
geostationary-transfer (GTO) orbits on a regular basis with the ESA Space Debris Telescope (ESASDT) located
on Tenerife, Spain. The aim of those surveys is to detect faint uncatalogued space debris objects. In these surveys declination stripes are
scanned in a so-called blind-tracking mode. In surveys of geostationary objects, the
telescope is kept fixed in starring mode, for surveys of objects in
geostationary-transfer orbits the apparent velocity of
reference objects is tracked. A survey of the ESASDT consists of 15 to 30 frames, spaced by one
minute. Also so-called targeted follow-up observations of known objects are
performed to improve the initial orbit. Those consist of up to 11 frames,
spaced by 30 seconds. In between exposures, the telescope is repositioned on
the same star field. The ESASDT has a field of view of 0.7$\times$0.7
degrees, resulting in two to three images per object in surveys and four to
six images per object in follow-up observations. In both cases, the
observations cover a time interval of two to three minutes.\\
Although a
priori information of the target object is available in follow-up
observations, the frames are processed with the same algorithms than the
surveys, in order to also detect further objects visible on the frames and to
link the object images of the target object in cases its orbital information
is very poor. \\
\\
Possibly many
object images are displayed on the single frames of a survey observation series. In a first step traces of cosmic ray events are
filtered on the single frames. In a second step, object images of the same
object on the different frames of the series are linked together. The linked
tracklets allow to extract the astrometric positions, which may lead to a
first orbit determination in further processing steps.

\section{Cosmic Filter and Object Image Link Algorithms}
\subsection{Cosmic Filters}
If charged particles hit a charged coupled device (CCD) detector photo-electrons are released punctually \cite{handbookCCD}. When the detector is read out, a cosmic ray impact leaves a trace similar to the one of photons. Not all charged particles hitting the detector are actually cosmic rays, but may also originate from weakly radioactive materials used in the construction of the CCD dewar, see e.g. \cite{Nielson}. In the current work all charged particles resolving photo-electrons at the detector are subsumed under the term \textit{cosmics}. \\
\\
A standard procedure in dealing with cosmics in astronomic imaging is, to
stack several images. The cosmics are filtered out because they are
\textit{singular} events irregularly spread over the CCD frame. This
technique has the additional side effect that the signal to noise ratio of the
observed object is increased \cite{handbookCCD}. But stacking requires a
precise alignment of the frames relative to the observed objects. For a
successful stacking the images have to be aligned relative to the object's
motion, an alignment with the stars would filter out the object images of fast
moving objects. In surveys, when searching for new objects, the object's
position on the frame and their exact motion are unknown. Therefore, possible
ranges of object velocities and inclinations have to be assumed and many
different stackings of the same frames have to be checked and the results
cross-checked. Such an approach is suggested and implemented by T.\,
Yanagisawa at the Mount Nyukasa observatory of the Japanese Aerospace
Exploration Agency (JAXA) \cite{toshiSD09}. This method is computationally
intensive and time consuming and not (yet) feasible for real-time
processing. Real-time processing is required, when tasking follow-up
observations within the same night. Additionally, stacking poses limits on the
object's orbits and its apparent movement at a very early processing step
already. Therefore, in the ESASDT processing a filter approach is preferred on
the single frames for cosmic rejection. A cosmic filter should be fast, and
allow to distinguish cosmics from actual object images. A filter approach acting on
single frames is only feasible if the pixel scale is small enough to allow for
such a discrimination, i.e., when the object images and cosmics are displayed
over a couple of pixels on the frame, to allow to determine distinguishing
features. \\
\\
Four different approaches have been used:\\
A so-called contrast filter has been developed, which filters cosmics via a threshold value for the ratio
between intensity value of the brightest pixel $\text{i}_{\text{peak}}$ and and
the mean intensity of the four surrounding ones $\text{i}_{\text{sur}}$,
calibrated with the background intensity $\rho$:
\begin{eqnarray}
&\text{threshold}=\frac{\text{i}_{\text{peak}}-\text{noise}\cdot\sqrt{\sigma_1}-\rho}{0.5(\text{i}_{\text{sur}}+\text{noise}\cdot\sqrt{\sigma_2})-\rho}
\end{eqnarray}
with
\begin{eqnarray}
&\sigma_{1}=\sigma^2_{\rho}+\text{gain}\cdot\text{max}(\text{i}_{\text{peak}},\rho)\\
&\sigma_2=\sigma^2_{\rho}+\text{gain}\cdot\text{max}(\text{i}_{\text{sur}},\rho)
\end{eqnarray}
with $\sigma_{\rho}$ being the standard
deviation of the improved background intensity, and $gain$ being the value for
the camera gain. The threshold value is determined empirically.\\
\\
A so-called object image class filter is based on six threshold values. Those
threshold values are determined empirically for five different object image
classes. The threshold values are combined equally weighted. An object image class is defined as the number of pixels an object image
candidate has on the frame:
\begin{eqnarray}
&\text{threshold}_1=\text{i}_{\text{peak}}-\rho \hspace{0.5cm}
&\text{threshold}_2=\text{i}_{\text{all}}-\rho\\
&\text{threshold}_3=\frac{\text{i}_{\text{all}}}{\text{i}_{\text{peak}}}\hspace{1.2cm}
&\text{threshold}_4=\frac{\text{FWHM}_x}{\text{FWHM}_y}\\
&\text{threshold}_5=\frac{\text{FWHM}_x}{n_{\text{pix}}}\hspace{0.5cm}
&\text{threshold}_6=\frac{\text{FWHM}_y}{n_{\text{pix}}}
\end{eqnarray}
whereas, $\text{i}_{\text{peak}}$ is the intensity of the brightest pixel,
 $\text{i}_{\text{all}}$ the mean of the intensity of all pixel of the object
 image, $\rho$ the improved background intensity, $\text{FWHM}_x$ and
 $\text{FWHM}_y$ an approximation of the FWHM in x- and y-direction with
respect to the pixel coordinates, and $n_{\text{pix}}$ the number of pixel
belonging to the candidate image on the frame. The FWHM is approximated by
counting the number of pixels in x- and y-direction of the pixel coordinates,
which are above the mean intensity of the whole object image in relation to
all pixels of the object image. This approximative calculation of the FWHM is
preferred over a Gaussian fit \cite{kouprianov}. The computational load
of a Gaussian fit is higher, which affects real-time processing, but more
importantly, a Gaussian fit presupposes an approximately Gaussian or Gaussian
elongated brightness distribution of the object images on the frames. This
assumption leads to wrong results for highly distorted images or for images
partly below signal to noise level.\\
\\
Two classical noise sensitive edge detection filters were used, Sobel and
Prewitt filters. To ease the computational burden and since the candidate
images on the frames are already detected, not the complete frame is filtered
with the kernels but only the smallest pixel-box containing the object image,
padded with a two pixel sized background border. The kernel for the Sobel filter are \cite{sobelsobel}:
\begin{eqnarray}
&\mathbf{G_x}=\begin{pmatrix} 1 & 0 & -1\\ 2 & 0 & -2 \\ 1 & 0 & -1 \end{pmatrix}\hspace{2cm}
\mathbf{G_y}=\begin{pmatrix} 1 & 2 & 1\\ 0 & 0 & 0 \\ -1 & -2 & -1 \end{pmatrix}\\
&\mathbf{G}=\sqrt{\mathbf{G_x}²+\mathbf{G_y}²}
\end{eqnarray}
The kernel of the Prewitt filter are \cite{prewitt}:
\begin{eqnarray}
&\mathbf{G_x}=\begin{pmatrix} -1 & 0 & 1\\ -1 & 0 & 1 \\ -1 & 0 & 1 \end{pmatrix}\hspace{2cm}
\mathbf{G_y}=\begin{pmatrix} -1 & -1 & -1\\ 0 & 0 & 0 \\ 1 & 1 & 1 \end{pmatrix}\\
&\mathbf{G}=\sqrt{\mathbf{G_x}²+\mathbf{G_y}²}
\end{eqnarray}
The gradients in the x- and y-direction are summed up. \\
\\
All filters have been
tuned conservatively using ten nights of the ESASDT observation campaigns
between January and July 2006.

\subsection{Object Image Linking}
\begin{figure}
\centering 
\includegraphics[height=1.2cm]{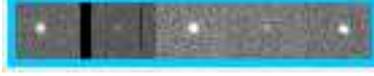}
 \caption{Five images spaced by 30 seconds of the same object observed with the ESASDT in January 2006.}
  \label{brightnessvariation}
\end{figure}

\begin{figure}
\centering         
\includegraphics[height=3.5cm]{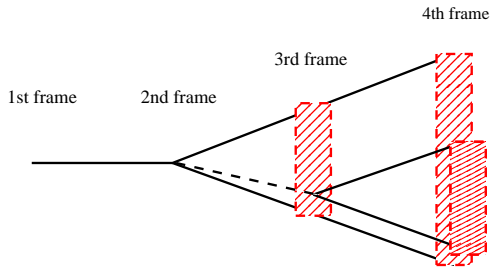}
 \caption{Linking object images on four frames. The boxes mark the allowed
   region.}
  \label{corr}
\end{figure}

Even after filtering, possibly many different object image candidates (true
object images of many different objects and left over cosmics)  are displayed and it has to be decided, which object images in fact
belong to the same object. This process is called linking of the object images
and positions, respectively. The appearance of the object on the single frames
can vary heavily, as Fig.\ref{brightnessvariation} shows, in which the same
object is displayed on five images of one series, spaced by 30 seconds only. Single object images can be missing although the object was in the field of view during the exposure, in the following called \textit{gap}, because belong signal to noise level or in front of a star image. A correlation of the single unlinked object images
with a catalogue, as done e.g. by \cite{fabrizioIAC07},  is not recommended:
Firstly, the correlation itself can be much improved in taking apparent velocity information into
account \cite{ichSD09}, especially when dealing with optical observations. Secondly,
no unknown objects cannot be determined from unlinked images only.  \\
\\
With the current observation strategy of the ESASDT, as explained in the
introduction, two to three object images per object in GEO and GTO survey
observations spaced by one minute, and four to six object images per object in
follow-up observation series spaced by 30 seconds are available. In both cases
the observations cover a time interval two to three minutes per object. A linear
accelerated movement of the objects over the frames during the short
observation series is assumed in J2000.0 space-fixed topocentric reference system. Two object image candidates on two subsequent
frames are preliminarily linked and their apparent movement is calculated in
standard coordinates in a tangent plane on the (topocentric) celestial sphere
\cite{sphere}. This has the advantage to be independent of the specific
observation scenario. This movement is
used to define so-called allowed regions, in which the next object image is
expected. An allowed region is a confidence region around the point of the expected next object image position in the tangent plane. The allowed region is defined by a deviation limit in the absolute value of the velocity
 and the direction of the velocity determined from the first two preliminarily
 linked object images to a possible third object image. If no object image is found in the allowed region, a gap can be
tolerated. The allowed region is broadened up on the subsequent frame
necessarily. If an object image is found in the allowed
region, it is added to the preliminary linked tracklet and the apparent
velocity is updated, see Fig.\ref{corr}. \\
\\
In a next step, so-called cross linking conflicts are
resolved: Cross-links are object images, which are linked in more than
one tracklet. In a first step, all cross-links are allowed, and it is noted to
which of possibly many different tracklets the single images belong. In a
second step, the length of the tracklets, cross-linked by the same object
image, are compared. The tracklet with the largest length, that is, the one
containing of the most object images, keeps, the cross-linked image, all
others are dissolved into single unlinked object images again. If two
tracklets with the same length are cross-linked, the similarity of the
cross-linked image with the preceding and following image within each tracklet
is evaluated: The values of the main axis of inertial tensors of the preceding
and following image are subtracted from the corresponding ones of the cross-linked image. The
tracklet with the larger differences is dissolved. The procedure is iterated until no cross-links occur any more. The remaining
images are checked for new links of the newly dissolved images again. In a
final step, links are made, in cases, if only two object images are available:
They are linked if they fall within an
apparent velocity and pseudo-inclination limit and if their inertial tensor fulfill a
similarity criterion. The link via pseudo-inclination and apparent drift is
necessary, since sometimes only two object images are available.\\
To determine the apparent drift $D$, and pseudo-inclination $I$ limit, the raw pixel coordinates are transformed into right ascension/declination ($\alpha/\delta$) and azimuth/elevation ($az/el$) using the pointing direction of the telescope:
\begin{equation}
D=\frac{\sqrt{(az_2-az_1)^2\cos^2(el_1)+(el_2-el_1)^2}}{(t_2-t_1)}
\end{equation}
\begin{equation}
I=\arctan{\Big(\frac{\delta_2-\delta_1}{(\alpha_2-\alpha_1)\cos(\delta_1)}\Big)}
\end{equation}
Currently a limit for the drift of 15 arcseconds per second and an inclination limit of
25 degrees is used. This allows to detect objects in a GEO region but puts
significant limits on detectability of all other orbits. To link only two
images of objects in other orbits, different drift and inclination intervals have to be chosen. The values of the main axis of the
inertial tensor of the two images are allowed to deviate by less than 20
percent. The latter criterion is only applied if both images are brighter than
800 ADU, to avoid misinterpreting the inertial tensor of faint disintegrated
traces close to the signal to noise level.\\
\\
40 manually selected tracklets, which were linked only via the apparent
drift and inclination limit and were manually corrected or were linked by hand, were evaluated in
order to find empirical limit values for the deviations in absolute velocity and direction of velocity. The tracklets
where picked randomly from the ESASDT campaigns of the first months of 2006. Only tracklets containing many object images were selected; the average was 4.5
images per tracklet. The absolute value of the apparent velocity
calculated from the first two object images be $|v_0|$ and the absolute value
of the apparent velocities calculated from the subsequent object image pair
within the set be $|v_i|$, the difference in the direction of the velocity
vector in the tangent plane be $\sphericalangle v$; the expectation value and standard deviation of the
velocity limit are determined as the following:
\begin{align}
\big<|v_i|/|v_0|\big>&=0.185\%\hspace{0.5cm}\sigma_{|v_i|/|v_0|}=0.156\%\label{exdrift}
\end{align}
\begin{equation}\begin{split}
\big<\sphericalangle v\big>&=0.001\mathrm{rad}\hspace{0.1cm}\hat{=}\hspace{0.1cm}0.063\mathrm{deg}
\\
\sigma_{\sphericalangle v}&=0.001\mathrm{rad}\hspace{0.1cm}\hat{=}\hspace{0.1cm}0.063\mathrm{deg}
\label{exwink}
\end{split}\end{equation}

\section{Performance with Real Observations}
The frames of the ESASDT are $2\times2$
binned in order to reduce the minimum signal-to-noise level for detectable
objects. At the observation site of the ESASDT the seeing is normally larger
than the pixel scale, therefore a binning is advantageous. The cosmic filters have been tested with ten survey observation series and
three follow-up series of the night of August 25, 2006. The results are
summarized in Table\,\ref{cosi}. In the appendix a few examples are shown of the detected object
image candidates and the ones still present after filtering by the different algorithms. \\
\setlength{\arrayrulewidth}{0.5pt}
\setlength{\doublerulesep}{0.6pt} 
\begin{table}
\begin{center}
\begin{tabular}{llllll}
 &\bf{Truth} &Contrast&Obj. Class&Sobel&Prewitt\\
\hline
\small object images&\bf{96}&84&85&84&79\\
\small cosmics &\bf{1527}&679&321&917&889\\
\small obj. rate&\bf{100}\%&87.5\%&88.5\%&87.5\%&82.2\%\\
\small cosmic rate&\bf{0}\%&55.5\%&79.0\%&36.4\%&41.8\%\\
\hline
\vspace{-0.3cm}\\
\end{tabular}
\caption{Number of detected object images and cosmics on ten observation
  series of the ESASDT taken on August 25, 2006: Truth and filtered results.}
\label{cosi}
\end{center}
\end{table}\\
The performance of rejecting the most cosmics is best for the contrast and
object class filter. All filters misidentify true object images as cosmics,
although they have been tuned as conservatively as possible. The Sobel filter
is slightly superior to the Prewitt filter in the tested setup. But not all filters do misinterpret the same object images as
cosmics, as a few examples listed in the appendix clearly show: The object
class filter misinterprets the bright and large object images with more than 30 pixels (e.g. Fig.\,\ref{30}), or object small object
images with very few pixels, with a high peak intensity (e.g. Fig.\,\ref{10A}
and \ref{26}). The latter cases could be as well cosmics as real object
images, by eye inspection they were judged to be more likely object images. The contrast filter algorithm misinterprets mostly the object images with a small number of pixels, similar to the object class filter, as e.g. Fig.\,\ref{27A}  and \ref{26} show. The filter also misinterprets object images with medium pixel size, see, e.g., Fig.\,\ref{10A}. Here, the peak intensity of this object image is relatively high, but it clearly is an object image. \\
\\
The performance of the linking algorithm is tested on two nights of the Tenerife campaign 2006 (January 26 and August 25) were used as an example. Table\,\ref{comp} shows the results. \\
\begin{table}
\begin{center}
\begin{tabular}{ll}
\hline
sets&781\\
\hline
correct sets&158\\
\hline
correct sets $\geq$3im&121\\
\hline
wrong sets $\geq$3im&2\\
\hline
images per correct set&4.19\\
\hline
manually corrected&0\\
\hline
high inclination&5\\
\hline
\end{tabular}
\caption{Tracklet linking of two nights of the Tenerife Campaign 2006 where analyzed with in total 117 observation series.}
\label{comp}
\end{center}
\end{table}\\
The link of two object images is prone to mistakes. If more than two object
images are available, the link is reliable. Only, two wrong sets have been
linked with more than two object images. With those tracklets it was possible
with both sets to determine a first orbit, but judging the single images from eyesight, it was decided that they seem to show cosmics. No
manual corrections needed to be applied to the sets itself, no images of
different objects have been linked and no cosmics have been embedded in a set
of more than two real object images. Restrictions on the orbit of link-able
objects are weak enough to allow to link the images of objects, which are in high inclination orbits.

\section{Discussion of Methods}
\subsection{Limitations of Possibilities by the Cosmic Filter}
None of the cosmic filters is perfect when the aim is not to loose any object
images. Filters which are based on many empirical parameters are highly
dependent on the specific camera and have to be carefully re-tuned with every
hardware change. In the step of linking object images it has to be assumed
that a considerable amount of cosmics is still present.

\subsection{Random Links}
In the presence of left over cosmics after filtering and also possible many
different object images displayed on the single frames, the probability of
random (wrong) links with the current algorithm needs to be
investigated. Observation series are assumed consisting of single frames containing a number of $q$ candidate
object images. The probability of random links of tracklets consisting of $m$
candidates is investigated. The allowed number of gaps is assumed to be $n$. A
square field of view is assumed. The probability for a random candidate
linking, i.e., the probability that tracklets are detected, although no object
is displayed on the frames, is estimated: For the estimation the candidates of
a whole observation series are added up in a squared single frame, called summary
frame in the following, with a given field
of view (FOV). The density of candidates within this summary frame is
calculated. The size of all allowed regions for a tracklet with a given number of images and gaps is determined under the
premise that the whole tracklet (including gaps) can be displayed in the
summary frame. This is in
accordance with the ESASDT observation strategy. The probability to find a
candidate in the allowed region with the given density of candidates is
evaluated, which leads to the following expression:
\begin{align}\begin{split}
&P\leq\sum\limits_{i=0}^{n}\Big(\frac{8}{(i+m-1)^2}
\cdot(|v_i|/|v_0|)_{\mathrm{limit}}\cdot \sin(\sphericalangle v_{\mathrm{limit}}/2)\cdot q \cdot (i+1)\Big)^{m-2}
\label{gleichung}
\end{split}\end{align}
for $m\geq 3$.\\
The probability is slightly overestimated, especially in long observation
series, since the density of candidates on the single observation frames is
significantly lower than on the summary frame.
\begin{figure}  
\centering
\includegraphics[width=0.5\textwidth]{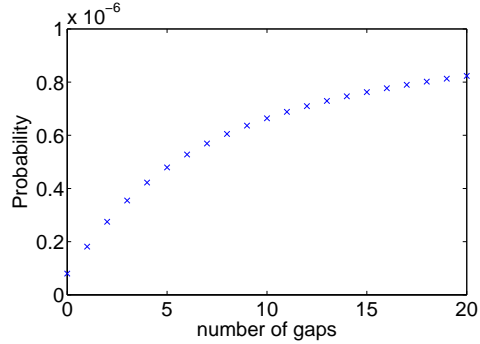}
\hspace{1.0cm}
 \caption{Probability of random links of tracklets of four candidate object images as a function of the number of allowed gaps within the set.}
  \label{plot1}
\end{figure}
\begin{figure}
\centering
\includegraphics[width=0.5\textwidth]{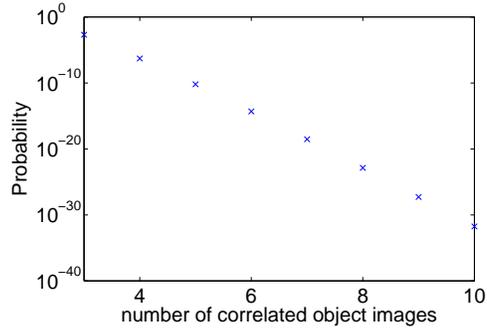} 
 \caption{Probability of random linking as a function of the number of candidate object images in the tracklet (logarithmic scale).}
  \label{plot2}
\end{figure}
\begin{figure}
\centering
\includegraphics[width=0.5\textwidth]{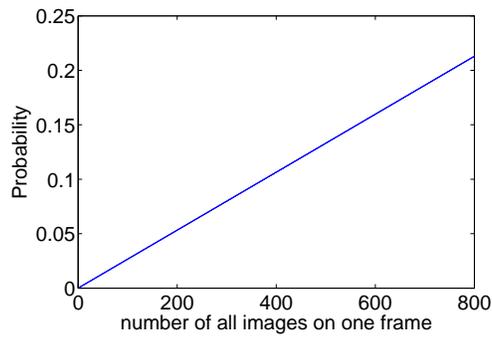} 
 \caption{Probability of random links as a function of the number of candidate
   object images on one frame, (a) for linking of three candidates with a
   maximum number of two allowed gaps.}
  \label{plot33}
\end{figure}
\begin{figure}
\centering
\includegraphics[width=0.5\textwidth]{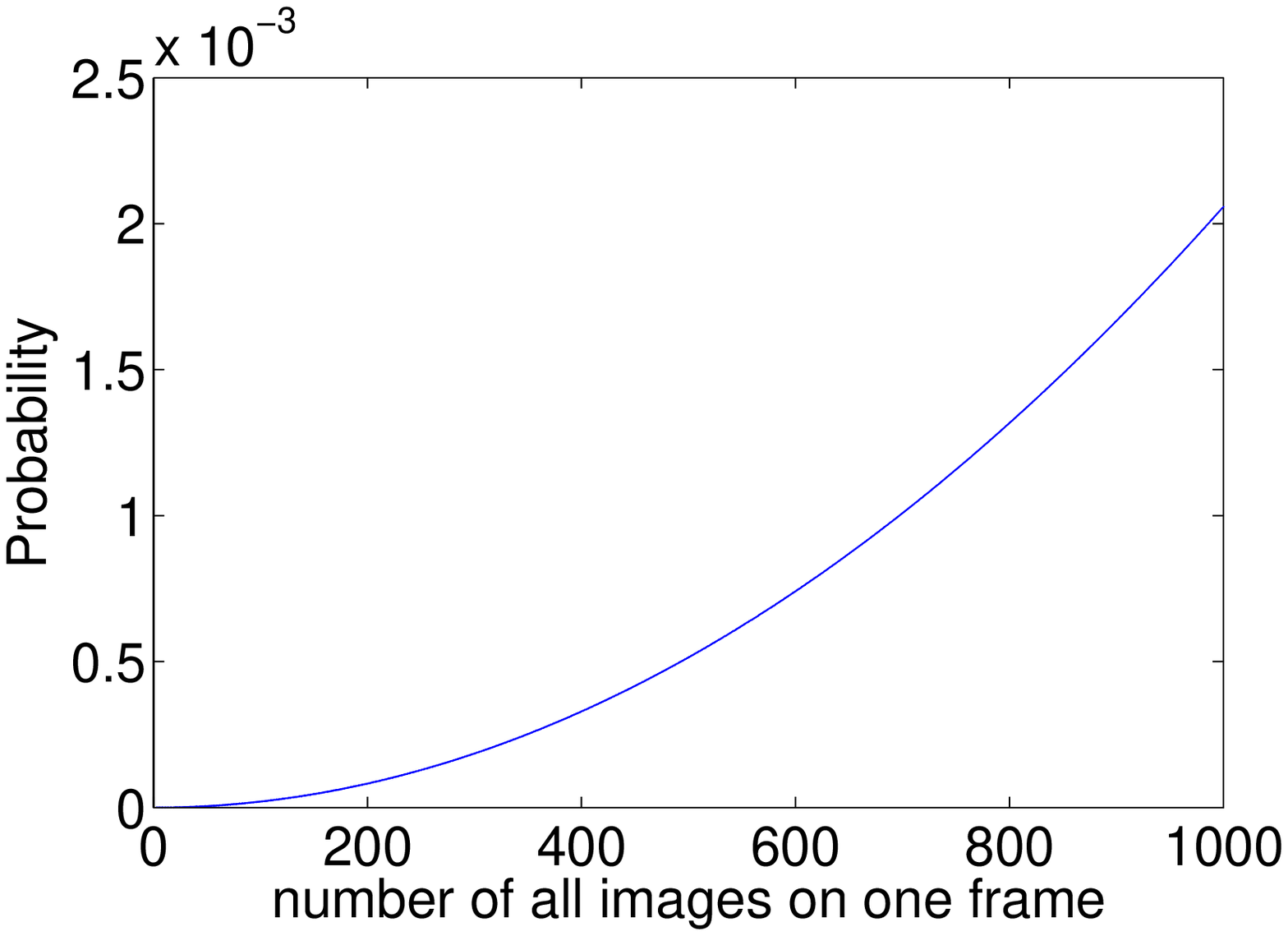}
 \caption{Probability of random links as a function of the number of candidate object images on one frame for linking of four candidates with a maximum number of six gaps.}
  \label{plot3}
\end{figure}
\\
If 30 candidates are assumed on every frame, the probability of random links
is $3.12\cdot10^{-3}$, if three gaps are allowed. If 300 candidates are
assumed on every frame the probability is
$2.52\cdot10^{-2}$, if two gaps are allowed. If only one gap is allowed the probability already
decreased to $1.80\cdot10^{-3}$ even for 300 object images on each frame. If
only one more object images is linked, so four in total, the
probability of random links decreases by more than one order of magnitude. Figure\,\ref{plot1}
shows the probability for linking four candidates into a tracklet as a function
of the number of gaps allowed for 30 candidates per frame. The probability
grows slowly with the number of allowed gaps. Figure\,\ref{plot2} shows the probability of random links as a
function of the number of candidates within the tracklet. The probability of
random links rapidly decreases with every candidate added to the
tracklet. Figure\,\ref{plot33} and \ref{plot3} shows the crucial dependence on
the total number of candidates on each frame for tracklets with three or four
candidates with a maximum of two and six allowed gaps, respectively. The
probability for random links of candidates is generally higher for tracklets
with only three candidates. The probability for random links grows linear for
tracklets of the candidates and quadratically for tracklets with four
candidates with the number of candidates images on the frames. \\

\subsection{Limits Imposed by Assumption of Steady Linear Apparent Motion}
\setlength{\arrayrulewidth}{0.5pt}
\setlength{\doublerulesep}{0.7pt} 
\begin{table}
\begin{center}
\begin{tabular}{p{2.3cm}p{2.3cm}llll}
\small COSPAR &\small Incl. (deg)& Ecc.& Tracklets\\
\hline\hline
\small\bf GEO& & &\\
\hline
05049B & 0.3184 & 0.0001 & 7\\
90102A & 12.8753 & 0.0002 &8\\
91010F & 12.2318  & 0.0015 &8\\
\hline
\small \bf GTO& & &\\
\hline
94056A & 14.9243  & 0.5024 & 88\\
97046D & 14.5544  & 0.4790& 87\\
\hline
95062C & 3.1615 & 0.8199& 90\\
99040D & 33.4316  & 0.8098& 91\\
\hline
\small \bf GPS& & &\\
\hline
84097A & 62.2247 & 0.0110& 19\\
85093A & 62.9986 & 0.0170& 21\\
\hline
\small \bf LEO& & &\\
\hline
58002B & 34.2430 & 0.1848& 414\\
\hline
\end{tabular}
\caption{Test objects and Number of Tracklets within 24 hours.}
\label{linear_objects}
\end{center}
\end{table}  
\begin{figure}
\centering
 \subfloat[\scriptsize{}]{\includegraphics[width=0.4\textwidth]{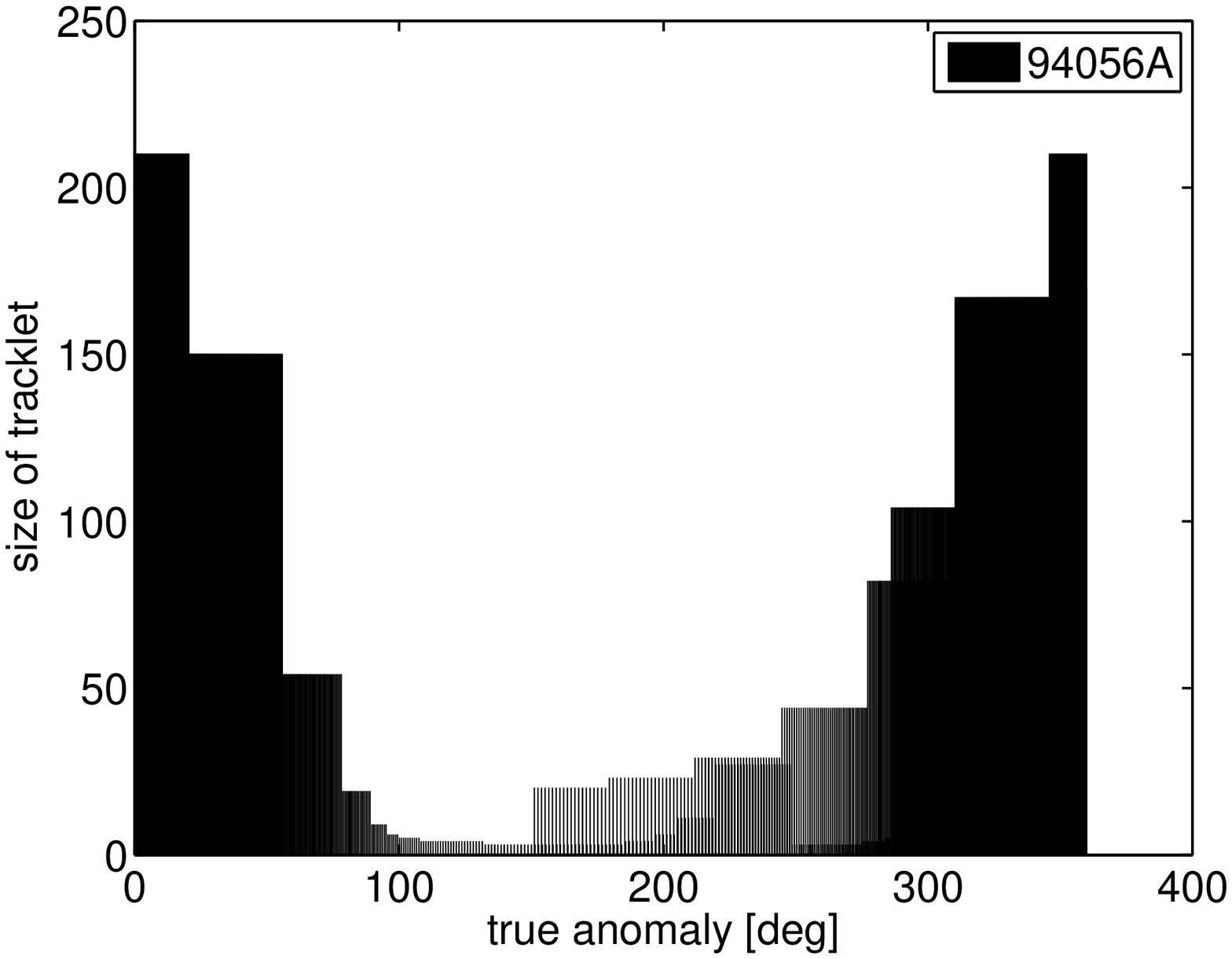}}   
 \subfloat[\scriptsize{}]{\includegraphics[width=0.4\textwidth]{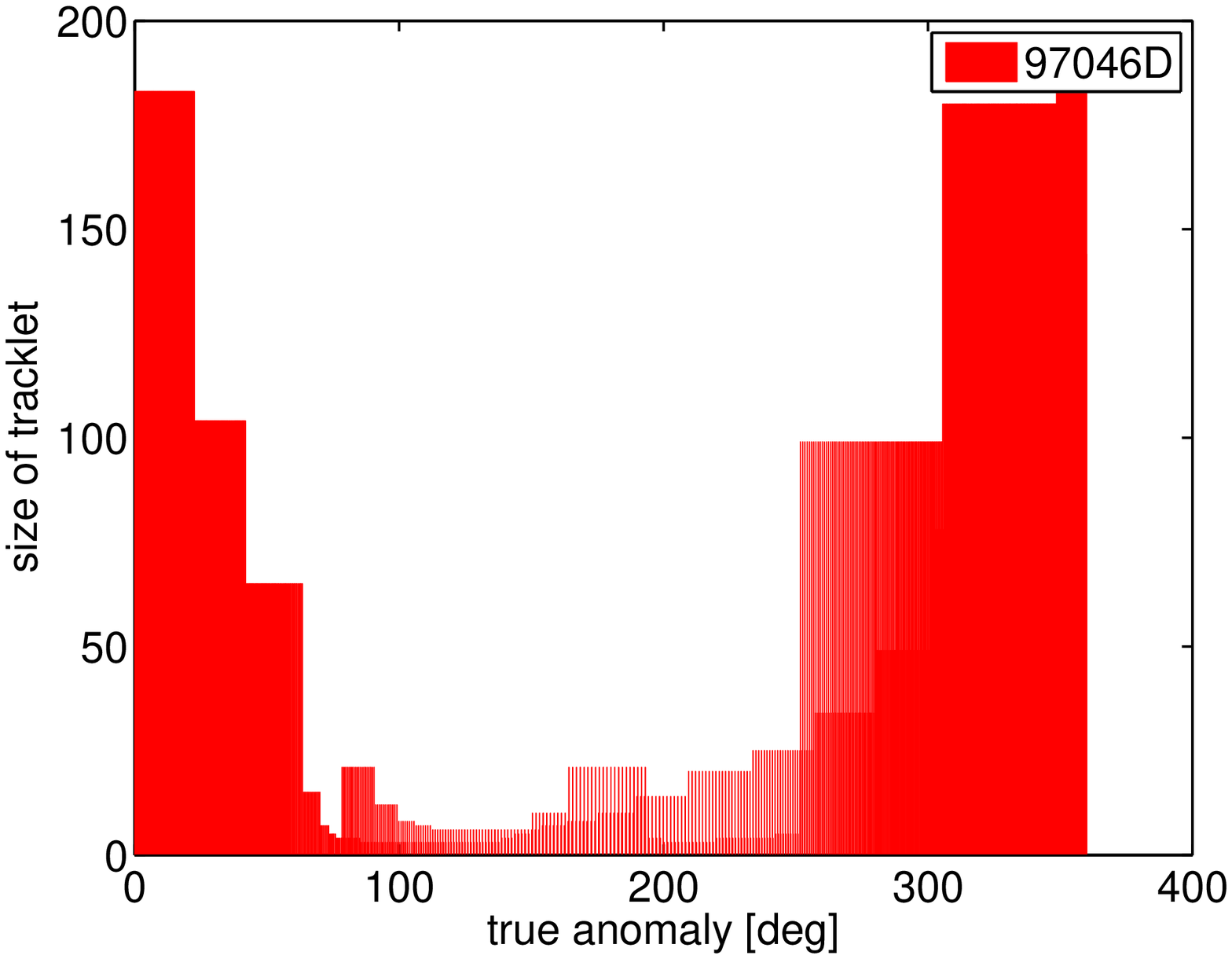}}
 \\
 \subfloat[\scriptsize{}]{\includegraphics[width=0.4\textwidth]{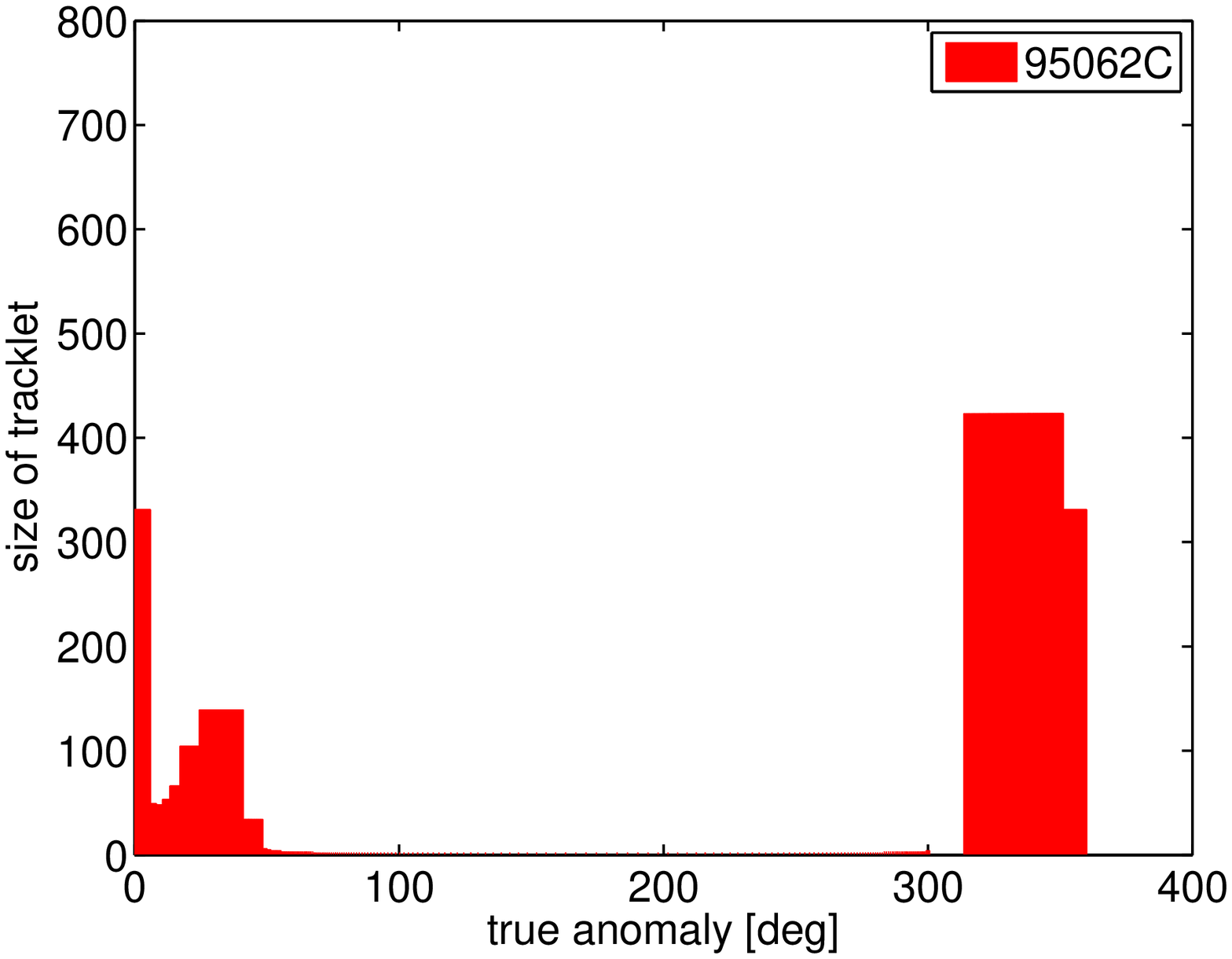}}   
 \subfloat[\scriptsize{}]{\includegraphics[width=0.4\textwidth]{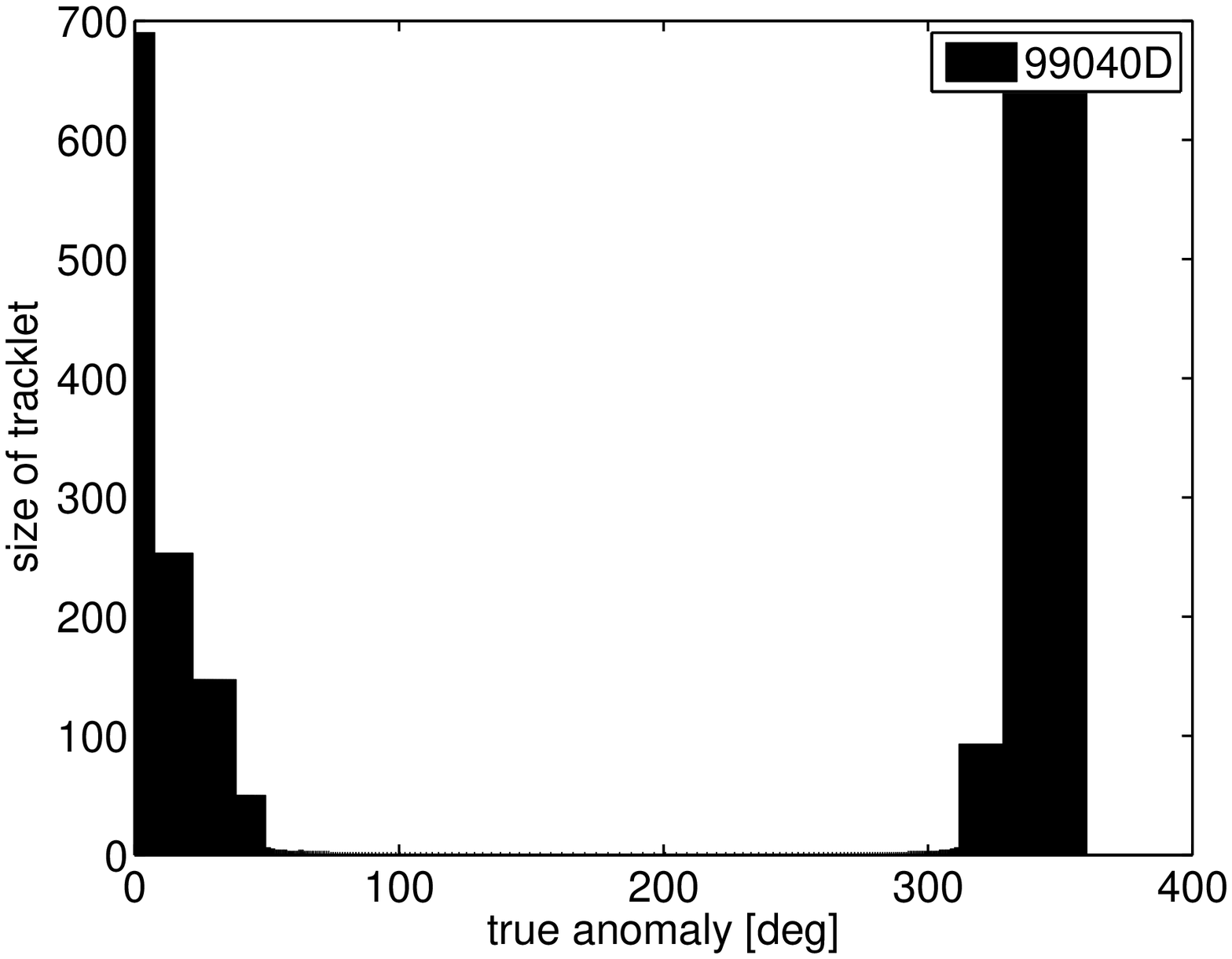}}\\
 \caption{Length of the tracklets in minutes and amount of single images
   respectively, as a function of true anomaly of the object
   (a) 94056A, (b) 97046D, (c) 95062C and (d) 99040D .}
  \label{anoGTO}
\end{figure}
The algorithm links object images based on the assumption of constant linear
apparent motion of the objects over the frames during the observation series,
but the velocity vector is updated with every successful link. Several orbits
were analyzed to study the limitations imposed by this assumption, they are
listed in Table \ref{linear_objects}. For all objects, geocentric ephemerides
are determined over a time interval of 24 hours with a spacing of one
minute. The ephemerides were transformed to the topocentric position of the
ESASDT. It was assumed that all objects are visible during 24 hours, i.e.,
that the earth is transparent. This assumption has been made so the linking
process is not limited by visibility constraints, but by the algorithm only.
The topocentric ephemerides are transformed in
so-called standard coordinates. A projection center has to be chosen for the
transformation. In a real observation scenario this center is given by the
pointing of the telescope. In the simulation, the topocentric viewing
direction of the first ephemerides position is chosen. An unlimited field of
view is assumed in the test setup: no new pointing is enforced by the object
running out of the field of view. The theoretical telescope is assumed to be
repositioned only, each time the algorithm is not able to connect the next
object image to the current tracklet (because the linearity condition is
violated, limit of the method is reached). After repositioning the object
position is again in the center of the tangent plane. Only linking in forward
direction is performed. If no third object image can be linked, the first two
object images remain linked and are counted as one tracklet in the
following. For the GEO objects not more eight tracklets have been formed over
24h hours, for the objects in different inclination orbits. For the objects in
high eccentricity orbits, 88 and 95 tracklets have been formed. But the length
of the tracklets is highly dependent on the anomaly, as Fig.\ref{anoGTO}
shows. For observations around the apogee the length of the tracklets is of
the same order as for GEO objects. For the GTO objects with moderate
eccentricities all tracklets with anomalies of over 300 and below 50 degrees
are longer than 80 minutes. For the two objects in high eccentricity orbits
tracklets longer than 150 minutes occur for anomalies larger than 320 and
below 30 degrees. For the GPS satellites about 20 tracklets have been formed, due to the small eccentricities no anomaly dependency occurs. For the LEO satellite 414 tracklets have been formed. The algorithm would have to be adjusted to a curved movement to be suitable for the LEO regime.\\

\section{Conclusions}
Both algorithms are implemented and used in the routine processing of GEO and
GTO surveys of the ESASDT, in order to detect new space debris objects. The
following concluding remarks can be made:\\
\\
It is possible to filter cosmics on the single observation
        frames. Classical edge detection filters appear to be inferior to
        empirical models in the detection of small pixel sized object images
        close to the signal to noise level, but deliver good results for
        bright large pixel object images. Empirical contrast filters deliver
        good results for faint object images.\\
\\
Assuming a linear steady accelerated motion for the movement of
        the objects over the frames without further constraints allows to
        reliably link object images. As long as more than two object images
        are available even with the tolerance of missing object images, and in
        the presence of cosmics and several hundred of object image candidates,
        reliable linking with a false rate of below 0.5 percent is possible.\\
\\
The linear accelerated motion, which is updated with every linked object
        image is a valuable approach in the link of images of GEO, and MEO
        objects. GTO object images are linked well as well near the
        apogee. For linking in the perigee region and for LEOs a curved
        reference motion is suggested.

\section*{Acknowledgments}
This work was supported by the Swiss National Science Foundation through grants 200020-109527 and 200020-122070 and the observations from the ESASDT were acquired under ESA/ESOC contracts 15836/01/D/HK and 17835/03/D/HK.

\bibliographystyle{authordate1}
\bibliography{frueh_short} 

\begin{appendix} 
\section{Examples of Cosmic Filter Results}
The results of the different cosmic filters are shown for some observation
series of the night August 25, 2006 of the ESASDT. First all detected
object image candidates are displayed in the observation series. In parenthesis first the number of real
object images judged by eyesight supported by object image linking algorithms
is displayed, secondly the overall number of detected candidates. For the
different filters first the number of true object images which are still
present after filtering is listed and the total number of candidates after
filtering. The true object images are marked with a box.
\begin{figure*}[h!]
  \centering
  \subfloat[no filter]{\includegraphics[width=0.7\textwidth]{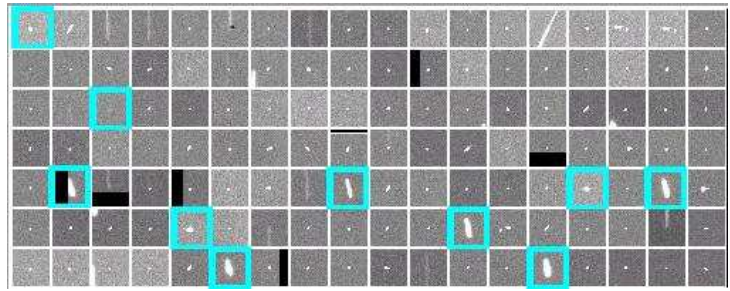}}\\
  \subfloat[contrast filter]{\includegraphics[width=0.7\textwidth]{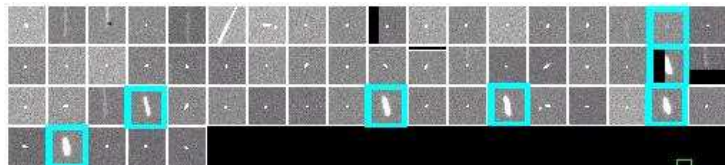}}\\
  \subfloat[object class filter]{\includegraphics[width=0.7\textwidth]{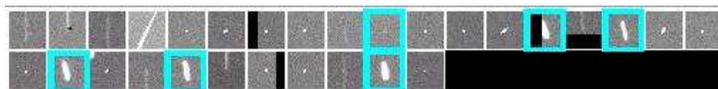}}\\
  \subfloat[Sobel filter]{\includegraphics[width=0.7\textwidth]{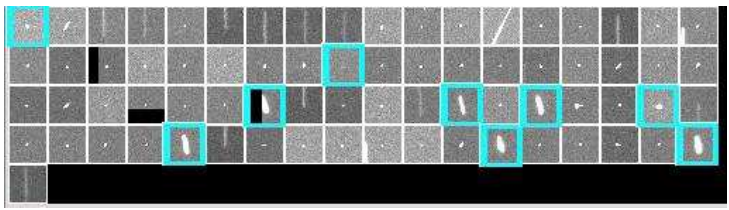}}\\
  \subfloat[Prewitt filter]{\includegraphics[width=0.7\textwidth]{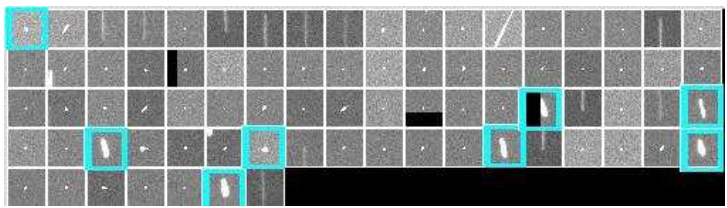}}
  \caption{(a) All detected candidates (10/116), (b) contrast filter (7/52), (c) object class filter (6/23), (d) Sobel (9/64), and (e) Prewitt (8/71) edge detection filter: SSO-10A.}
  \label{10A}
\end{figure*}\\
\begin{figure*}
  \centering
  \subfloat[no filter]{\includegraphics[width=0.7\textwidth]{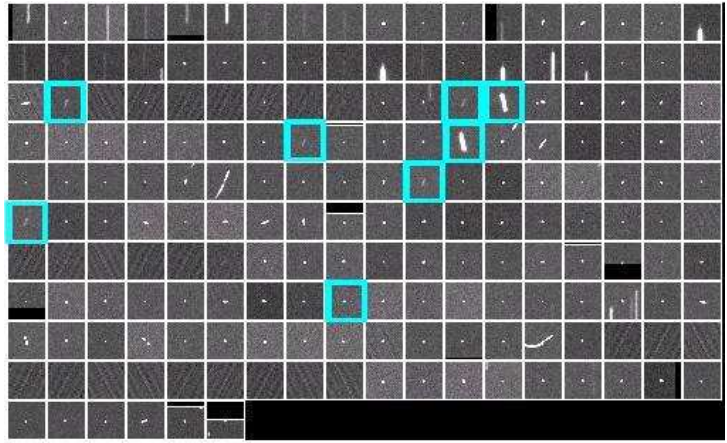}}\\
  \subfloat[contrast filter]{\includegraphics[width=0.7\textwidth]{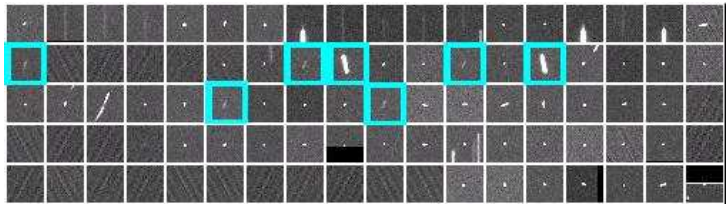}}\\
  \subfloat[object class filter]{\includegraphics[width=0.7\textwidth]{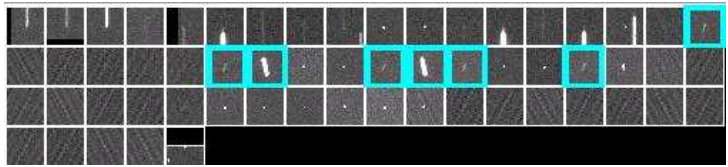}}\\
  \subfloat[Sobel filter]{\includegraphics[width=0.7\textwidth]{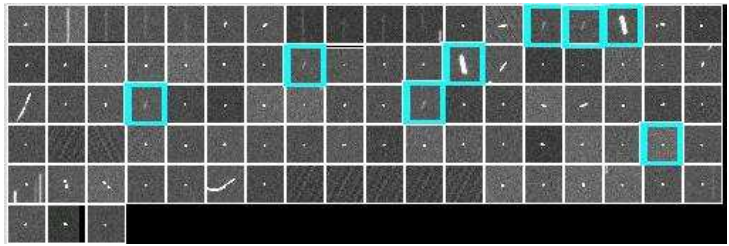}}\\
  \subfloat[Prewitt filter]{\includegraphics[width=0.7\textwidth]{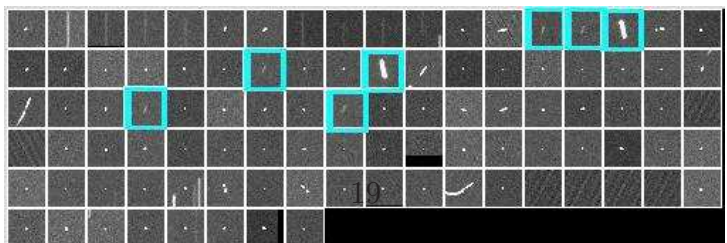}}
  \caption{(a) All detected candidates (8/178),(b) contrast filter (7/83), (c) object class filter (7/52), (d) Sobel (8/85), and (e) Prewitt (7/91) edge detection filter: SSO-27A.}
  \label{27A}
\end{figure*}
\begin{figure*}
  \centering
  \subfloat[no filter]{\includegraphics[width=0.7\textwidth]{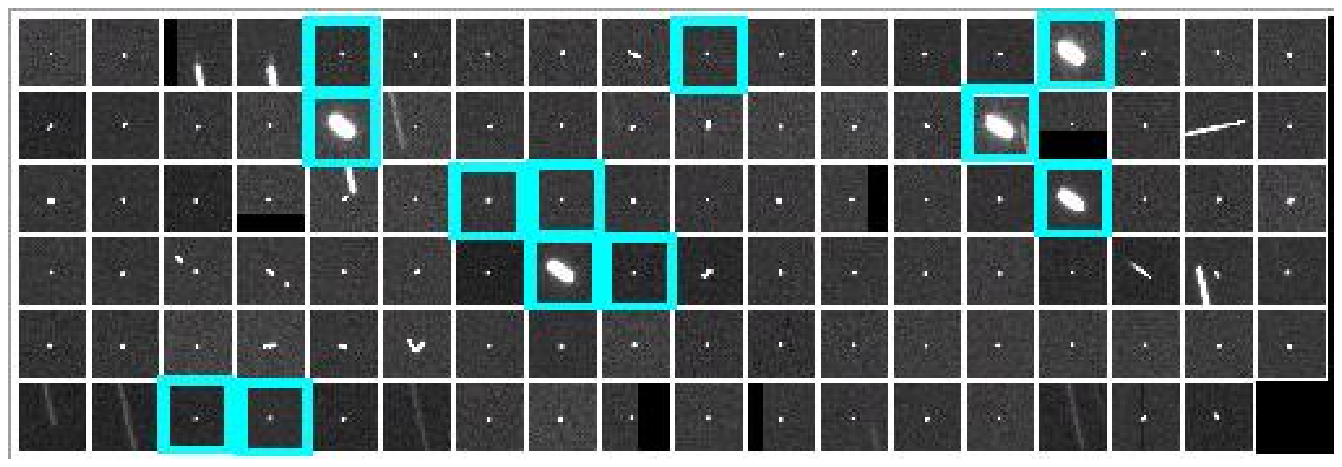}}\\
  \subfloat[contrast filter]{\includegraphics[width=0.7\textwidth]{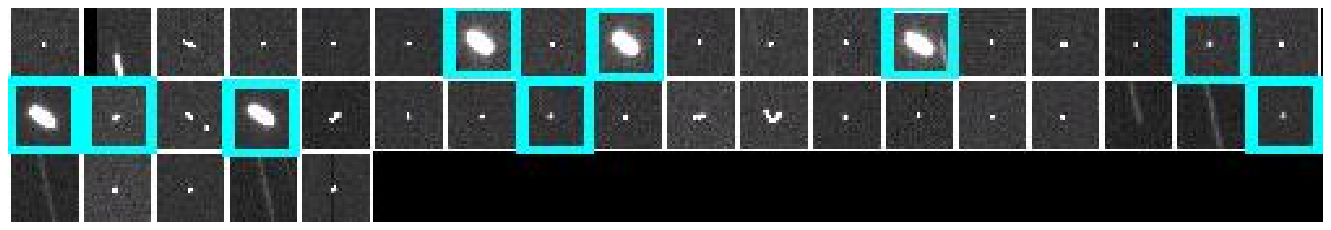}}\\
  \subfloat[object class filter]{\includegraphics[width=0.7\textwidth]{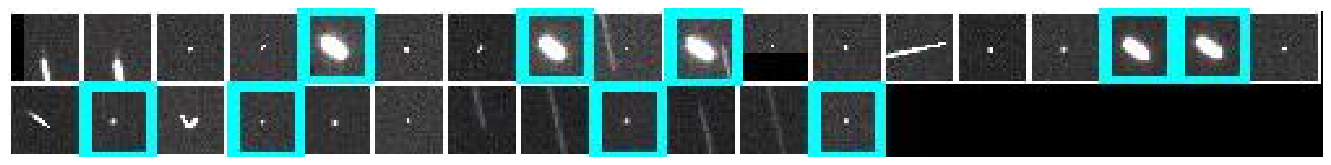}}\\
  \subfloat[Sobel filter]{\includegraphics[width=0.7\textwidth]{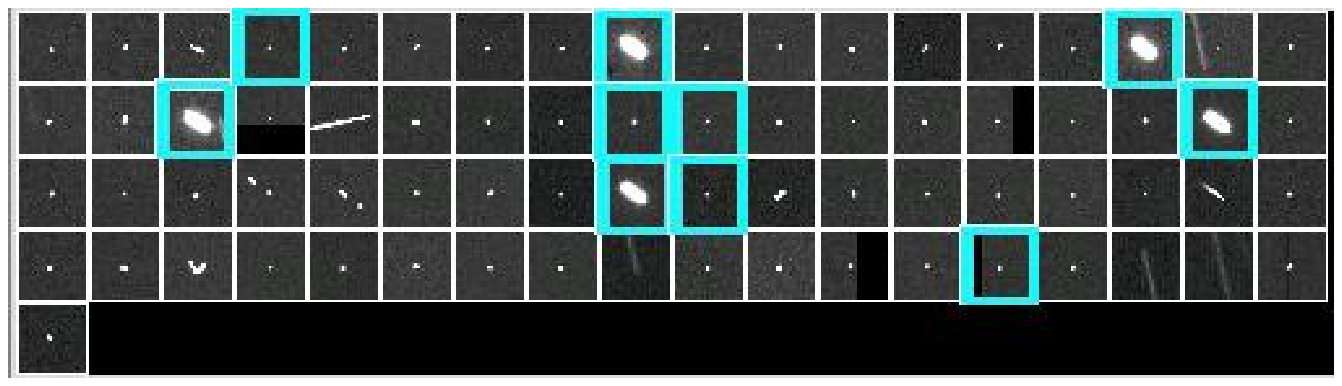}}\\
  \subfloat[Prewitt filter]{\includegraphics[width=0.7\textwidth]{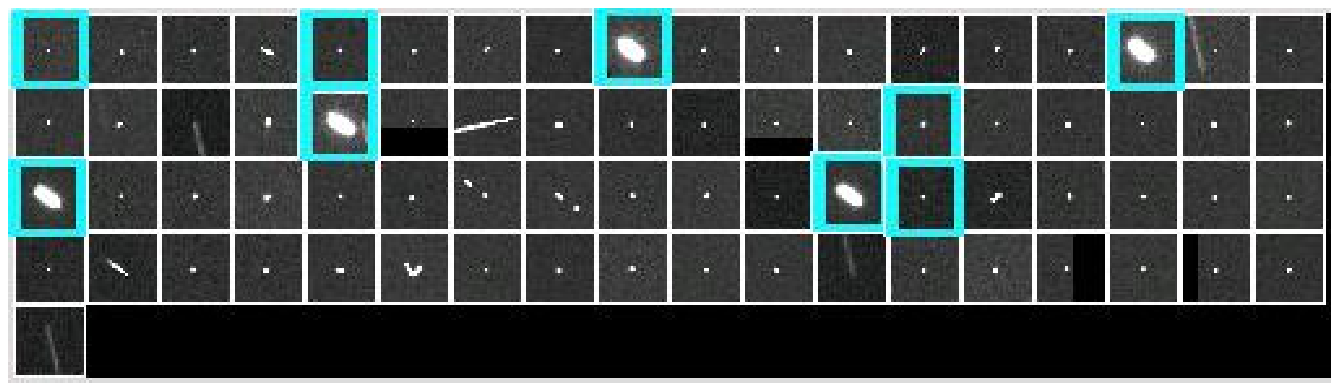}}
  \caption{(a) All detected candidates (12/95),(b) contrast filter (9/32), (c) object class filter (9/21)), (d) Sobel (9/63)), and (e) Prewitt (9/64) edge detection filter: SSO-26.}
  \label{26}
\end{figure*}
\begin{figure*}
  \centering
  \subfloat[no filter]{\includegraphics[width=0.7\textwidth]{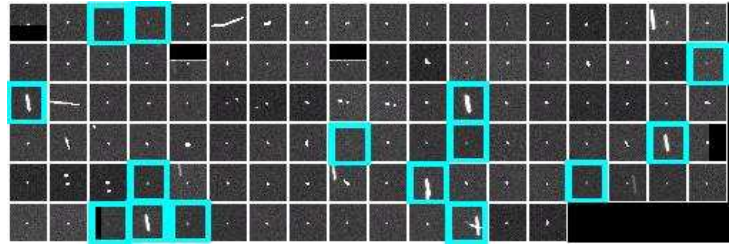}}\\
  \subfloat[contrast filter]{\includegraphics[width=0.7\textwidth]{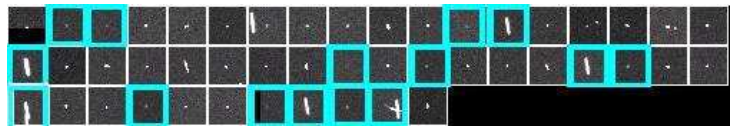}}\\
  \subfloat[object class filter]{\includegraphics[width=0.7\textwidth]{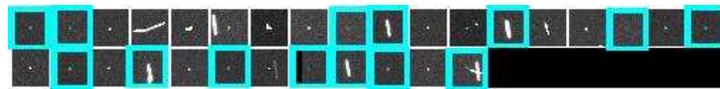}}\\
  \subfloat[Sobel filter]{\includegraphics[width=0.7\textwidth]{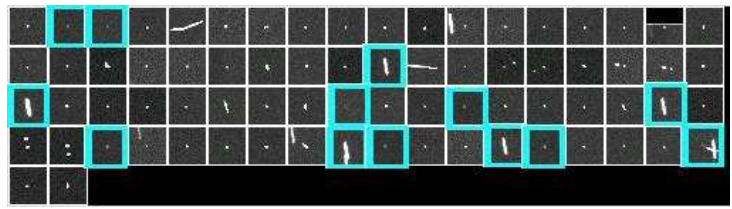}}\\
  \subfloat[Prewitt filter]{\includegraphics[width=0.7\textwidth]{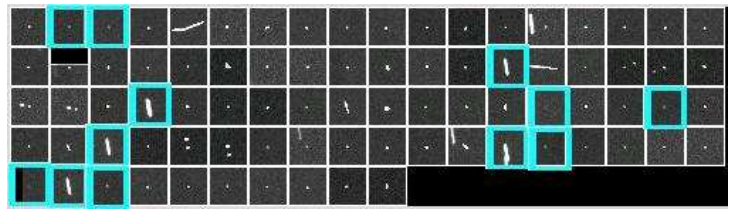}}
  \caption{(a) All detected candidates (15/89),(b) contrast filter (15/32), (c) object class filter (14/16), (d) Sobel (13/61), and (e) Prewitt (13/69) edge detection filter: SSO-30.}
  \label{30}
\end{figure*}
\end{appendix} 
\end{document}